\documentclass[twocolumn,nopacs,preprintnumbers]{revtex4-2}
\usepackage{graphicx}
\usepackage[latin1]{inputenc}
\usepackage{bm}
\usepackage[dvipsnames]{xcolor}
\usepackage[normalem]{ulem}

\usepackage{amsmath,amssymb}
\usepackage{hyperref}

\usepackage{physics}
\usepackage{subfigure}

\newcommand{\fin}{\text{f}}
\newcommand{\ini}{\text{i}}
\newcommand{\inter}{\text{int}}
\newcommand{\refe}{\text{r}}

\graphicspath{%
    {converted_graphics/}
    {/}
}
\begin{document}

\title{Taming the time evolution in overdamped systems:  shortcuts elaborated from  fast-forward and time-reversed protocols
}


\author{Carlos A. Plata$^{1,2}$, Antonio Prados$^2$, Emmanuel Trizac$^1$ and David Gu\'ery-Odelin$^3$}
\affiliation{
$^1$ Universit\'e Paris-Saclay, CNRS, LPTMS, 91405, Orsay, France \\
$^2$ F\'isica Te\'orica, Universidad de Sevilla, Apartado de Correos 1065, E-41080 Sevilla, Spain \\
$ ^ 3 $ Laboratoire Collisions, Agr\'egats, R\'eactivit\'e, IRSAMC, Universit\'e de Toulouse, CNRS, UPS, France 
}


\date{\today}

\begin{abstract}
Using a reverse-engineering approach on the time-distorted solution in a reference potential, we work out the external driving potential to be applied to a Brownian system in order to  slow or accelerate the dynamics, or even to invert the arrow of time. By welding a direct and time-reversed evolution towards a well chosen common intermediate state, we derive analytically a smooth protocol to connect two \emph{arbitrary} states in an arbitrarily short amount of time. Not only does the reverse-engineering approach proposed in this Letter contain the current---rather limited---catalogue of explicit protocols but it also provides a systematic strategy to build the connection between arbitrary states with a physically admissible driving. Optimization and further generalizations are also discussed.

\end{abstract}

\maketitle

Shortcut To Adiabaticity techniques originally aim at reaching adiabatic outcomes in a finite amount of time \cite{reviewSTA}. Adiabatic should be understood here  in its quantum sense, synonymous with ``sufficiently slowly driven''~\footnote{And not in the thermodynamic sense of zero heat.}. These methods have been  extended to generate shortcuts between two states, regardless of the existence of an adiabatic connection between them \cite{reviewSTA}. While this field is rooted in quantum mechanics~\cite{Chen10,Chen10b}, related questions emerge in other domains, namely classical mechanics \cite{Jarzynski13,Deng13,DGO14a,Deffner14,Kolodrubetz17,Gonzales17,Patra17} and stochastic thermodynamics \cite{Martinez16,Li17,Martinez17,chupeau18,rondin17,Patra17,dago2020,Funo20}. Yet, the question of transposing to statistical physics protocols originally developed in quantum mechanics is delicate. For instance, the so-called counterdiabatic protocol (also dubbed transitionless tracking) valid for any initial condition \cite{Demirplak03,Berry09} can be directly transposed to overdamped  dynamics \cite{Li17}. However, the very same procedure yields non conservative forcings, experimentally problematic to achieve, with underdamped systems \cite{Li17,reviewSTA}.  Among the set of tools for accelerating the dynamics, other classes of solutions propose tailor-made protocols based on the specifics of the initial and final states, both in quantum mechanics~\cite{Masuda08,Berry09,DGO14a,Vitanov15,davidFF,Kang16,Qi17} and in statistical physics~\cite{Schmiedl07,Martinez16,NJP18,Nakamura20}. More general optimal protocols in small thermodynamic systems  have been obtained from a mapping to optimal transport, establishing an unexpected connection with cosmology but requiring a numerical resolution~\cite{Aurell11,Aurell12,Aurell12b,Muratore17,Zh19}. These protocols generically lead to discontinuous-in-time driving forces~\cite{Schmiedl07,Aurell12}, which raises a delicate experimental challenge for implementation.
\begin{figure}[tbp]
\begin{center}
\includegraphics[width=0.28\textwidth]{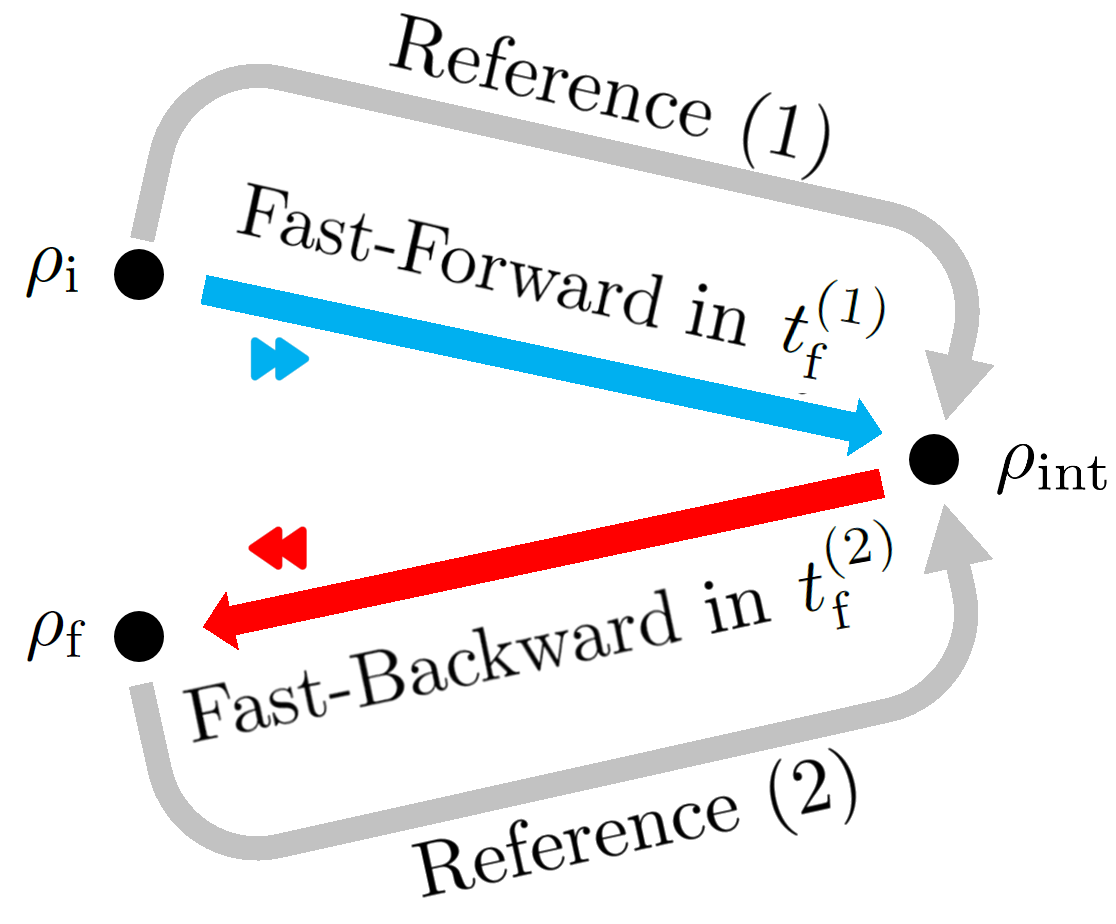}
\end{center}
\caption{
Sketch of the welding strategy to connect two arbitrary states, $\rho_{\ini}$ and $\rho_{\fin}$, through an auxiliary intermediate  $\rho_{\inter}$. The reference processes are the direct relaxations toward $\rho_{\inter}$ starting either from $\rho_{\ini}$ or $\rho_{\fin}$. An acceleration of the first reference process connects $\rho_{\ini}$ to $\rho_{\inter}$ in a time $t_{\fin}^{(1)}$. The second reference process is time reversed and accelerated; it lasts $t_{\fin}^{(2)}$. Combining the two steps allows to reach the target state $\rho_{\fin}$ in a final time $t_{\fin}= t_{\fin}^{(1)}+t_{\fin}^{(2)}$, providing a fast shortcut from $\rho_\ini$ to $\rho_\fin$. In general, $\rho_{\inter}$ differs from both $\rho_\ini$ and $\rho_\fin$. Otherwise, one of the steps (forward or backward) becomes unnecessary.}
\label{fig:gen-strategy}
\end{figure}

A central question is thus to work out shortcuts to transformations between arbitrary states, that should involve non-singular forces  and be expressible in closed form. This is the question we solve in the present Letter, within the framework of the Fokker-Planck equation \cite{risken1996}, which  governs the evolution of the probability density $\rho(x,t)$ of Brownian objects, with $x$ and $t$ denoting position and time.  To this end, we proceed in two steps. First, we show how to distort  and control a \textit{reference} dynamics, i.e., the time evolution taking place in a reference external potential $U_\refe(x,t)$, with a well chosen external potential $U(x,t)$. The realization of such drivings is now achieved experimentally in a number of domains, from cold atom physics \cite{davidlivre} to colloids \cite{Ciliberto17}, thanks to a proper steering of the optical potential applied to the system \cite{rondin17,dago2020,Albay20}. This will allow us to accelerate the reference dynamics (fast-forward) \cite{Masuda08}, or to decelerate it (slow-forward), and even to freeze the evolution. Interestingly, this also opens the way to reverse time's arrow and devise a backward drive at arbitrary speed. Second, we combine a fast-forward drive {\em towards}  a suitably chosen intermediate state $\rho_{\inter}(x)$ with a  fast-backward protocol {\em from} $\rho_{\inter}(x)$, to get a whole family of smooth shortcuts, as depicted in Fig.~\ref{fig:gen-strategy}. In doing so, the system can be brought from a chosen state $\rho_{\ini}(x)$  to another chosen  state $\rho_{\fin}(x)$, in an arbitrary time. Both states can be equilibrium or, more generally, out-of-equilibrium \cite{Baldassarri20,Prados21}.

\textit{Reverse engineered potential.-} The Brownian objects considered, such as colloids in dilute conditions, are immersed in a thermal bath at temperature $T$ and submitted to an external force field stemming from a potential $U(x,t)$. The Fokker-Planck equation for the probability density $\rho(x,t)$ reads
\begin{equation}
\label{eq:U_FP}
\partial_{x}\left[\rho(x,t)\partial_{x}U(x,t)\right]=-\beta^{-1}\partial_{x}^{2}\rho(x,t)+\gamma\partial_{t}\rho(x,t),
\end{equation}
where $\gamma$ is the bath friction coefficient and $\beta=(k_B T)^{-1}$, with $k_B$ the Boltzmann's constant. A unique solution for $U(x,t)$ can be obtained by imposing a prescribed time evolution for the density $\rho(x,t)$, from an initial $\rho(x,0)=\rho_{\ini}(x)$ to a final $\rho(x,t_{\fin})=\rho_{\fin}(x)$ state~\footnote{For the moment, we focus our attention on single-step connections. The concept of the intermediate state will return when designing the welding strategy.}. Once $\rho(x,t)$, together with the operating time $t_\fin$, have been chosen, Eq.~\eqref{eq:U_FP} is nothing but a first order differential equation for the external force  $-\partial_x U(x,t)$ that should be applied to the system to achieve the appropriate driving. Assuming that the density current vanishes at the boundaries of the system, $x=x_b$, we get 
\begin{equation}
\label{eq:dxu}
\partial_x U(x,t)   =  -\beta^{-1} \partial_x \ln \rho(x,t) +  \frac{\gamma }{\rho(x,t)} \partial_t \int_{x_b}^x \!\!dx'  \rho(x',t).
\end{equation}
Unfortunately, except in a few cases, like that of a Gaussian distribution~\cite{Schmiedl07,Martinez16}, shape preserving evolutions \cite{Zh20}, or other very specific situations \cite{Zh20,Zh19}, Eq.~\eqref{eq:dxu} is impractical since it does not lead to an explicit closed-form potential $U(x,t)$~\cite{SM}. We thus seek an alternative route that provides a systematic approach to obtain admissible driving protocols in closed-form.

\textit{From Fast-Forward (FF) to Fast-Backward (FB).-} The idea is to take advantage of the knowledge of a reference non-trivial dynamics for $\rho(x,t)$ to distort its time evolution by finding the appropriate 
driving potential.  First, we show how a time contraction can be performed to force the system to reach equilibrium in a finite amount of time. Such a FF protocol realizes in finite time a process that takes an infinite amount of time, when the system is unperturbed. Beyond FF, FB protocols---and also slow forward or backward ones---can also be engineered. We subsequently explain under which conditions the driving force remains continuous for all times, a key requirement for practical implementation.

Consider a reference process,  i.e., a known solution $\rho_{\refe}(x,t)$ of the Fokker-Planck equation \eqref{eq:U_FP} in a confinement potential $U_{\refe}(x,t)$ over a time interval $0<t<t_{\refe}$, which we write as a continuity equation
\begin{subequations}\label{eq:rho0} 
\begin{align}
  \partial_t \rho_{\refe}(x,t)& =-\partial_x \left[\rho_{\refe}(x,t) v_{\refe}(x,t) \right], \\
  v_{\refe}(x,t)&\equiv- \gamma^{-1} \partial_x\left[U_{\refe}(x,t)  +\beta^{-1}  \ln \rho_{\refe}(x,t)\right].
\end{align}
\end{subequations}
We define the desired prescribed density as
\begin{equation}
\label{eq:presc_rho}
\rho(x,t)\equiv \rho_{\refe}(x,\Lambda(t))
\end{equation} 
through a time manipulation of the reference process, embedded in the function $\Lambda(t)$. The prescribed and the reference evolution go through the same states, but displayed at a different frame rate and/or time ordering. Indeed, depending on the choice for $\Lambda(t)$, one can play ``the movie'' at FF speed ($\dot{\Lambda}>1$), at slow motion ($0<\dot{\Lambda}<1$), pause it ($\dot{\Lambda}=0$), or even rewind it: either at slow backward motion ($-1<\dot{\Lambda}<0$) or at FB speed ($\dot{\Lambda}<-1$). In the latter case, $\dot{\Lambda}<0$, our procedure enables a new possibility: the design of a potential to invert time's arrow for an irreversible evolution obeying the Fokker-Planck equation.

Introducing Eqs.~\eqref{eq:rho0} and \eqref{eq:presc_rho} into Eq.~\eqref{eq:dxu}, and defining $\Delta U(x,t)=U(x,t)-U_{\refe}(x,\Lambda(t))$, which measures the departure from the reference potential, we get 
\begin{align}\label{eq:U(x,t)-generic-v2}
\partial_x \Delta  U(x,t) =&   \left(1-\dot
         \Lambda(t)\right) \gamma v_{\refe}(x,\Lambda(t)),
\end{align}
where the time dependence of the protocol is encapsulated in $\Lambda(t)$ and its derivative $\dot{\Lambda}(t)$. For simplicity, we restrict ourselves to $\Lambda(t)$ that are monotonic functions of time, either of FF ($\Lambda(0)=0$ and $ \Lambda(t_{\fin})=t_{\refe}$) or of FB ($\Lambda(0)=t_{\refe}$ and $ \Lambda(t_{\fin})=0$) type. The times $t_{\refe}$ and $t_{\fin}$ can be finite or infinite. Of particular relevance for experimental applications though is the shortcut of an infinite process (infinite $t_{\refe}$ but finite $t_{\fin}$),  e.g., for building irreversible nano heat engines \cite{Schmiedl08,Bo13,Tu14,Martinez15,Dechant15,Martinez16_spain,Nakamura20,Plata20,Tu21}.

For the sake of simplicity, we consider hereafter a static reference potential $U_{\refe}(x)$. Such a potential is convenient; it makes it possible to analytically obtain $\rho_r(x,t)$  through an expansion in eigenmodes, as shown below \footnote{Yet, 
Eq.~\eqref{eq:U(x,t)-generic-v2} remains valid for a time-dependent reference potential, should this choice be more appropriate for a specific situation}.
With $t_{\refe} \to \infty$,
the reference process is thus the relaxation to the equilibrium distribution for $U_{\refe}(x)$: $\rho_\refe \propto e^{-\beta U_\refe}$.  In order to accelerate such an everlasting dynamics, we impose $\lim_{t\to t_{\fin}^-} \Lambda(t)=+\infty$. For our purposes, it is adequate to use the family of functions
\begin{equation}\label{eq:lambda-family}
    \Lambda(t)=\tau f\!\left(\frac{t}{t_{\fin}}\right), \qquad  f(z)=\frac{z^2}{(1-z)^{\zeta}}\, , \quad \zeta>0,
\end{equation}
where $\tau$ is a characteristic time. The divergence of $\Lambda(t)$ at $t=t_{\fin}$ implies that $\lim_{t\to t_{\fin}^{-}}\dot\Lambda(t)=+\infty$, which suggests that $\Delta U(x,t)$ may diverge at the final time. However, the velocity field $v_{\refe}(x,\Lambda(t))$ vanishes at $t_{\fin}$ and thus the behavior of $\Delta U(x,t_{\fin}^{-})$ needs to be elucidated. 

\textit{Regularity of the driving potential.-} The family defined in Eq.~\eqref{eq:lambda-family} has the property $\dot{\Lambda}(0)=0$. This guarantees the continuity of the force at the initial time, as implied by Eq.~\eqref{eq:U(x,t)-generic-v2}.
We show now on general grounds that the force field remains continuous even at $t_\fin$, provided the reference potential $U_{\refe}(x)$ is confining---in the sense that there exists a well-defined equilibrium density. The proof uses the expansion in eigenmodes for the solution of the reference process, and a mapping to a quantum problem. Introducing $\psi(x,t)=\rho_{\refe}(x,t)e^{\beta U_{\refe}(x)/2}$  leads to the time-dependent Schr\"odinger equation $\partial_t \psi = -H \psi$ with Hamiltonian~\cite{footnote1},
\begin{equation}\label{eq:H-eff}
H = - \frac{1}{\gamma \beta} \partial_x^2 + \frac{1}{2\gamma} \left[ \frac{\beta U_{\refe}'(x)^2}{2} - U_{\refe}''(x) \right].
\end{equation}
The smallest eigenvalue, associated to the equilibrium distribution, is zero: $H e^{-\beta U_{\refe}(x)/2}=0$. The other eigenvalues of $H$ are positive, $0<\lambda_{1}<\lambda_{2}<\cdots $, implying that the corresponding modes decay exponentially in time. Specifically,  $H \varphi_{n}(x)=\lambda_{n}\varphi_{n}(x)$, where $\varphi_{n}(x)$ is the eigenfunction associated to the $n$-th eigenvalue~\footnote{The eigenfunctions are mutually orthogonal and normalized, $ \braket{\varphi_m}{\varphi_n}=\int_{\cal D} dx\, \varphi_m(x) \varphi_n(x)=\delta_{mn}$,  where ${\cal D}$ denotes the integration domain.}.

The expansion of the reference process in the eigenbasis reads
\begin{equation}
  \label{eq:FP-expansion}
\rho_{\refe}(x,t)=\frac{e^{-\beta U_{\refe}(x)}}{Z_{\refe}}+ e^{-\beta U_{\refe}(x)/2} \sum_{n=1}^{\infty} c_{n}\varphi_{n}(x)e^{-\lambda_{n}t}
\end{equation}
where $Z_{\refe}=\int_{\cal D} dx\, e^{-\beta U_{\refe}(x)}$ is the partition function and $c_n=\braket{\varphi_n}{\psi(t=0)}=\int_{\cal D} dx\, \varphi_n(x) \psi(x,0)$. We are interested in the limit $t\to t_{\fin}^{-}$, for which $\Lambda(t) \to +\infty$. From Eq.~\eqref{eq:FP-expansion} we deduce
\begin{align}
  \label{eq:FP-expansion-v2}
 \rho_{\refe}(x,\Lambda(t)) \sim 
\frac{e^{-\beta U_{\refe}(x)}}{Z_{\refe}}\left(1+
                       \tilde{c}_{1}\chi_{1}(x)e^{-\lambda_{1}\Lambda(t)}\right),
\end{align}
where $\tilde{c}_{n}\equiv Z_{\refe}c_{n}$ and $\chi_{n}(x)\equiv e^{\beta U_{\refe}(x)/2}\varphi_{n}(x)$. As a result, to the lowest order in $e^{-\lambda_{1}\Lambda(t)}$, we get
\begin{equation}
  \partial_x \ln \rho_{\refe}(x,\Lambda(t)) \sim -\beta
    \partial_x U_{\refe}(x)+\tilde{c}_{1} \partial_x \chi_{1}(x)e^{-\lambda_{1} \Lambda(t)},
\end{equation}
which combined with Eq.~\eqref{eq:U(x,t)-generic-v2} gives
\begin{equation}
\label{eq:DU_aymp_tf}
 \partial_x \Delta U(x,t)\sim -
  \beta^{-1}\frac{\tilde{c}_{1}}{\lambda_1}\frac{de^{-\lambda_{1}\Lambda(t)}}{dt}
  \partial_x \chi_{1}(x).
\end{equation}
There is a vast set of diverging functions $\Lambda(t)$ including the family \eqref{eq:lambda-family} that forces the cancellation of the right hand side of  Eq.~\eqref{eq:DU_aymp_tf} when $t$ approaches $t_{\fin}$ \cite{SM}, a requirement that implies that the driving force remains continuous at the final time. In our proof, the existence of a well-defined partition function $Z_{\refe}$ is important. This is not the case for a free expansion, $U_{\refe}(x)=0$, in an infinite or semi-infinite space \cite{SM}. However, the argument remains valid for a reference free diffusing system within a finite box, as shown below.

To sum up, Eq.~\eqref{eq:U(x,t)-generic-v2} provides the smooth potential needed to tune at will the reference process according to the time mapping function $\Lambda(t)$. Despite the limited number of analytically solvable reference processes~\cite{SM}, we can always rely on an expansion as in Eq.~\eqref{eq:FP-expansion}, but with a certain cutoff---examples are provided below. In doing so, the tailored process does not strictly  reach the target state, but its distance thereto---e.g., with the $L^2$ norm---can be made as small as desired~\cite{SM,carslaw_h_s_introduction_1950}.
\begin{figure*}
\begin{center}
\includegraphics[width=0.9\textwidth]{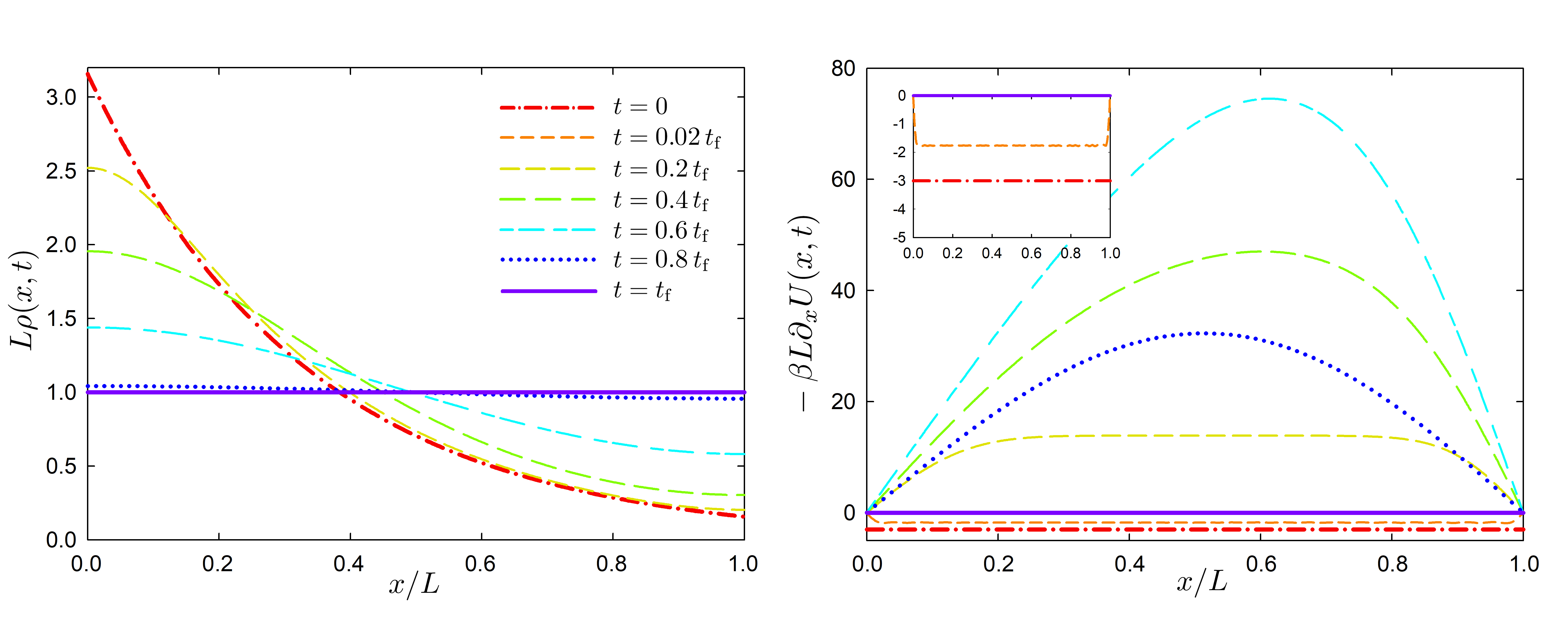}
\end{center}
\caption{Accelerated free diffusion in a finite-size box of size $L$. Starting from the exponential profile $\rho_{\refe}(x,0) \propto e^{\beta F x}$, the subsequent reference solution $\rho_{\refe}$ corresponds to a force-free evolution, $U_{\refe}(x)=0$. This leads to strict equilibrium in an infinite time, featuring a characteristic (diffusive) time $\tau$. This reference dynamics is accelerated such that at $t_{\fin}=\tau/10$, the density profile is strictly that at equilibrium (flat,  i.e., $\rho(x,t_\fin)=1/L$). The time evolution of the density is displayed in the left panel, whereas the right panel shows the  force required to drive such an accelerated transformation. The inset shows a zoomed region for negative forces. Here, $\beta F L =-3$ (as conventional for a gravitational field, we take $F<0$). For the numerical evaluation, the cutoff in the expansion \eqref{eq:FP-expansion} is  $n_{\text{cut}}=70$~\cite{SM}.
}
\label{fig:box}
\end{figure*}

\textit{Diffusion in a box.-} We illustrate the time manipulation procedure with the FF transformation of the free diffusion for a dilute gas within a finite box, $x \in [0,L]$. The gas is initially at equilibrium in the presence of a constant and homogeneous force, $\rho_{\refe}(x,0) \propto e^{\beta F x}$, with thus a sedimentation profile. In the reference process, at time $t=0$, the force is suddenly removed and it takes an infinite time to reach the stationary homogeneous state. The time scale of the relaxation is characterized by $\lambda_1^{-1}=\gamma \beta L^2 / \pi^2$. 

We have chosen $\Lambda(t)$ from Eq.~\eqref{eq:lambda-family} with $\zeta=1$, $\tau=\lambda_1^{-1}$, and $t_{\fin}=\tau/10$. The results for the FF transformation are displayed in Fig.~\ref{fig:box}, for the density of the gas
and the force $-\partial_x U(x,t)$. Apart from a short time window where the forces are negative because of continuity, the forces required to accelerate homogenization are positive,  pushing the system in the direction opposite to that of the initial force. As highlighted above, the protocol is smooth in time, including the initial and final times. In general, faster accelerations have a higher cost. Both the magnitude of the required force and the excess irreversible work increase as $t_{\fin}$ is decreased~\cite{SM}.

\begin{figure*}
\begin{center}
\includegraphics[width=0.9\textwidth]{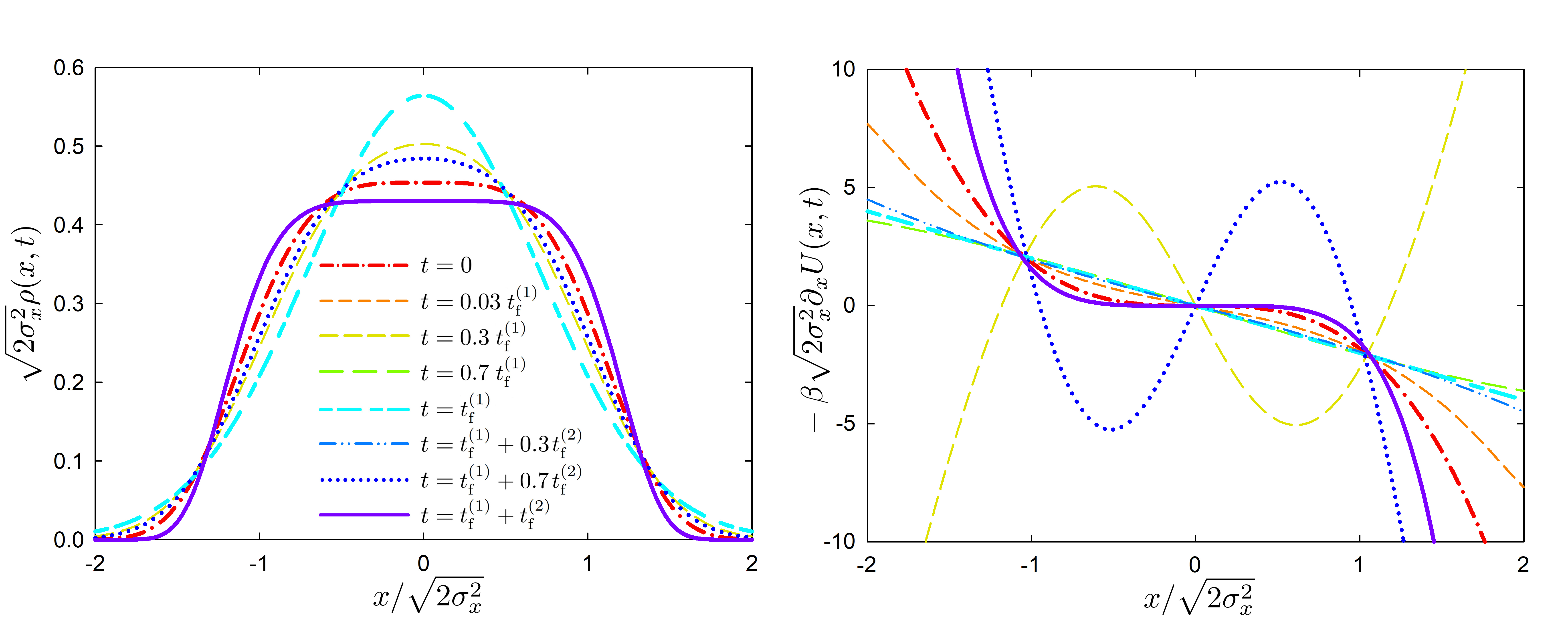}
\end{center}
\caption{Illustration of an operational welding connection. Left: time evolution of the distribution. Right: driving force to be applied. The intermediate state is Gaussian with variance $\sigma_x^2$, 
associated to the reference potential
$U_{\refe}(x)=x^2/(2\sigma_x^2)$. The initial (final) state is the thermal Boltzmann distribution in the potential $U_{\ini}(x)\propto x^4$ ($U_{\fin}(x) \propto x^6$).  The initial, final, and reference densities are chosen to have the same variance. The transformation is performed in a finite time $t_{\fin}=t_{\fin}^{(1)}+t_{\fin}^{(2)}$  with $t_{\fin}^{(1)}=t_{\fin}^{(2)}=\tau/10$; $n_{\text{cut}}=70$~\cite{SM}.
}
\label{fig:harm}
\end{figure*}

\textit{Connecting arbitrary states.-} We come back to the welding idea conveyed in Fig. \ref{fig:gen-strategy}. For the sake of concreteness, the initial and final densities are the equilibrium distributions corresponding, respectively, to the initial and final potential $U_{\ini}(x) \propto x^4$ and  $U_{\fin}(x) \propto x^6$.  None of these two potentials provides a convenient reference potential: the associated reference dynamics cannot be analytically solved and the FF idea results inoperative. Here,  the welding method offers a solution. Indeed, choosing  a Gaussian intermediate state ($\log\rho_\inter \propto -x^2$) is free of the above difficulties, and solvable. We thus take  $\beta U_{\refe}(x)=x^2/(2 \sigma_x^2)$,  where $\sigma_x^2$ stands for the variance of $\rho_\inter$ \cite{SM}. As previously, we accelerate the forward (FF) and backward (FB) processes by a factor ten, $t_{\fin}= t_{\fin}^{(1)}=t_{\fin}^{(2)}=\tau/10$, with $\tau=\lambda_1^{-1}=\gamma \beta \sigma_x^2$. For both the FF and FB steps, the $\Lambda$ function  is taken to be the same, with $\zeta=1$~\footnote{To account for time reversal and the time shift in the FB part, $t\to t_{\fin}^{(1)}+t_{\fin}^{(2)}-t$.}. The results for the welded FF and FB protocols are displayed in Fig.~\ref{fig:harm}. In the first step, the tails of the density have to be pushed away from the center and the confinement needs to be strengthened in the central region to reach faster $\rho_{\inter}$; hence the $N$-shape force. During the second stage, the requirements are opposite, leading to an inverted $N$-shape. The force protocol is continuous for all times, including the initial and final times for both steps of the welding strategy. 

We have focused on a specific functional shape for the time manipulation, Eq.~\eqref{eq:lambda-family},  in order to ensure the continuity properties that we desire for the protocol. One may consider optimization problems, such as finding the time manipulation that minimizes some relevant observable---like the excess irreversible work~\cite{Aurell11,Aurell12,Aurell12b,Muratore17,Zh19}.  Remarkably, it is possible to show that such an optimization over all FF protocols (attached to a given $U_\refe$ but with an arbitrary $\Lambda(t)$) lead to processes delivering excess work at constant rate, as it happens with the full optimization~\cite{SM}. Besides, when it comes to connecting two Gaussian states, the  optimal FF
protocol coincides with the full, unconstrained, optimum---and the minimum restricted to the specific $\Lambda(t)$ family in Eq.~\eqref{eq:lambda-family} lies only 4\% above it, for a ten-fold expansion of the state~\cite{SM}.

To sum up, we have developed a reverse-engineering technique in order to manipulate at will the time evolution of a reference process. This provides us with the external potential required to reach a target distribution in a desired time. Interestingly, not only does the framework allow for the acceleration of forward processes but also for the inversion of time's arrow. Taking these time manipulated reference processes as building blocks, we have put forward a neat welding procedure to connect two arbitrary states in an arbitrarily small finite time. The method relies on a non-linear time mapping of two relaxation processes in the same harmonic potential; it produces by construction a smooth driving potential, with continuous force field, for all times. Finally, our procedure can be generalized to higher dimensional problems\cite{2021Frim,SM}.  Some possible venues for developments lie in the thermalization processes to build irreversible nano heat engines \cite{Schmiedl08,Bo13,Tu14,Martinez15,Dechant15,Martinez16,Nakamura20,Plata20,Tu21} or the genetic control of an evolving population \cite{2021Iram,2021Weinreich}.

\begin{acknowledgments}
We would like to thank S. Ciliberto, S. Dago and L. Bellon for fruitful discussions. This work was supported by the Agence Nationale de la Recherche research funding Grant No.~ANR-18-CE30-0013, and project PGC2018-093998-B-I00 funded by FEDER/Ministerio de Ciencia e Innovaci\'on--Agencia Estatal de Investigaci\'on (Spain). C.A.P.  also acknowledges financial support from Junta de Andaluc\'{\i}a and European Social Fund through the program PAIDI-DOCTOR.
\end{acknowledgments}

\bibliography{FFbib}

\begin{thebibliography}{58}
\bibitem [{foo()}]{footnote1}%
  \BibitemOpen
  \href@noop {} {}\bibinfo {note} {The advantage of this formulation lies in
  the use of the usual definition of scalar product in Hilbert space, with a
  weight function equal to unity. Equivalently, the Fokker-Planck operator can
  also be made Hermitian by introducing a scalar product with weight function
  $e^{\beta U_{\refe}(x)}$}\BibitemShut {NoStop}%
\end{thebibliography}

\end{document}